\documentclass[11pt]{article}
\usepackage{amssymb}
\usepackage{amsmath}

\def\R{{\mathbb R}}
\def\Z{{\mathbb Z}}

\def\N{{\mathbb N}}

\def\F {{\mathcal F}}

\def\s {{\sigma}}

\newtheorem{theo}{Theorem}
\newtheorem{prop}{\indent Proposition}

\newtheorem{rem}{\indent Remark}

\newtheorem{ex}{\indent Example}
\newtheorem{ass}{\indent Assumption}

\title{Modeling networks of spiking neurons as interacting processes with memory of variable length}

\date{February 21, 2015}
\author{A.~Galves \and E.~L\"ocherbach}

\begin{document}

\maketitle

\begin{abstract}
We consider a new class of non Markovian processes with a countable number of interacting components, both in discrete and continuous time. Each component is represented by a point process indicating if it has a spike or not at a given time. The system evolves as follows. For each component, the rate (in continuous time) or the probability (in discrete time) of having a spike depends on the entire time evolution of the system since the last spike time of the component. In discrete time this class of systems extends in a non trivial way both Spitzer's interacting particle systems, which are Markovian, \nocite{Spitzer1970}  and Rissanen's stochastic chains with memory of variable length which have finite state space.\nocite{rissanen} In continuous time they can be seen as a kind of Rissanen's variable length memory version of the class of self-exciting point processes which are also called ``Hawkes processes'', however with infinitely many components. These features make this class a good candidate to describe the time evolution of networks of spiking neurons. In this article we present a critical reader's guide to recent papers dealing with this class of models, both in discrete and in continuous time. We briefly sketch results concerning perfect simulation and existence issues, de-correlation between successive interspike intervals, the longtime behavior of finite non-excited systems and propagation of chaos in mean field systems. 

\end{abstract}



\section{Introduction}
A biological neural system has the following characteristics. It is a system with a huge (about $10^{11}$) number of interacting components, the neurons. The activity of each neuron is represented by a point process, namely, the successive times at which the neurons emit an action potential or a so-called spike. It is generally considered that the spiking activity is the way the system encodes and transmits information. 

The spiking probability or rate of a given neuron depends on its membrane potential. 
The membrane potential of a given neuron is affected by the actions of all other neurons interacting with it. Neurons interact either by {\sl chemical}  or by {\sl electrical} synapses. Chemical synapses can be described as follows. Each neuron spikes randomly following a point process with rate depending on the membrane potential of the neuron.  At its spiking time, the membrane potential of the spiking neuron is reset to an equilibrium potential $0$. At the same time, simultaneously, the neurons affected by it receive an additional {\sl amount of potential} which is added to their membrane potential. 

Electrical synapses occur through {\sl gap-junctions} which allow
neurons in the brain to communicate with one another. This induces an
attraction between the values of the membrane potentials of each other
and, as a consequence, a drift of the system towards its center of
mass. Finally, leakage channels may induce a loss of membrane potential for each neuron. 

The fact that the membrane potential of each neuron is reset to $0$ when it spikes makes its time evolution to be dependent of a variable length of the past. More precisely, it depends on the influence received from its presynaptic neurons since its last spiking time. 
In other terms, the time evolution of such a system is obviously not described by a Markov process (Cessac 2011\nocite{cessac_discrete_time}). 
In particular, if we consider a time continuous description of the system, the waiting times between two successive spikes of a single neuron are not exponentially distributed (see, for instance, Brillinger 1988\nocite{Brillinger1988}). 

Such a system can be described in discrete time in the following way. Consider a small interval of time, typically of the order of $10 ms$ which is more or less the time it takes for a neuron to emit a spike, followed by a refractory period. We indicate the presence or absence of a spiking activity for each neuron within each such time window. Then the process we obtain is a system of interacting chains with memory of variable length and a large number of components. This class of  systems extends in a non trivial way both the interacting particle systems, which are Markov, see Spitzer (1970),\nocite{Spitzer1970}  and the stochastic chains with memory of variable length which have finite state space, see Rissanen (1983)\nocite{rissanen} or Galves and L\"ocherbach (2008)\nocite{galves-locherbach-rissanen}. 

A continuous time description of the process seems however more convenient both from a modeling and a mathematical point of view. This is the point of view adopted in De Masi et al. (2015), Duarte and Ost (2014) and Fournier and L\"ocherbach (2014). 

The present article is considered as a critical reader's guide to the papers mentioned above. We will briefly sketch the main results of these papers as well as challenges and next steps to be addressed. 

This paper is organized as follows. In Section \ref{section:def} we introduce a model of an infinite network of spiking neurons in discrete time; in Section \ref{sec:continuous time} an analogous continuous time model using point processes is introduced. In Section \ref{sec:Markov}, we show that under appropriate conditions, such a system of interacting point processes can be represented via an associated interacting particle system which is Markovian. In Section \ref{sec:results1} we give an existence and perfect simulation result for the process defined in Sections \ref{section:def} and \ref{sec:continuous time}. Section \ref{sec:results2} considers a finite system composed of $N$ neurons where the graph of synaptic weights is a realization of a (slightly) supercritical directed Erd\"os-R\'enyi random graph. In this case, the correlations of two neighboring interspike intervals are shown to be asymptotically de-correlated, as the system size tends to infinity. Sections \ref{sec:longtime} to \ref{sec:further results} are devoted to a study of the associated interacting particle system introduced in Section \ref{sec:Markov} where we deal successively with the longtime behavior of finite particle systems, with the hydrodynamical limit within a mean field system and finally with asymptotic properties of the limit process. In Section \ref{sec:questions} we mention challenges and next steps to be addressed. We close our paper with a discussion in Section \ref{sec:discussion}.

\section{Infinite systems of interacting processes with memory of variable length in discrete time }\label{section:def}
We introduce a new class of stochastic processes, both in discrete and in continuous time, which are models of networks of spiking neurons.  The processes we  consider are infinite systems of interacting processes with memory of variable length. 

Let $I$ be a countable set of neurons and introduce a family of  {\it synaptic weights} $W_{j \to i} \in  \R , $ for $j \neq i ,$ $W_{j \to j } = 0 $ for all $j.$  We interpret $W_{j\to i}$ as the {\it synaptic weight of neuron $j$ on neuron $i$.} We suppose that the synaptic weights have the following property of uniform summability
\begin{equation}\label{eq:summable}
\sup_{ i \in I} \sum_j |W_{j \to i }| < \infty .
\end{equation}
Moreover, we shall use a family of  {\it spiking rate functions} $\phi_i : \R  \to [ 0, 1 ], i \in I, $ and a family of  {\it leak functions} $ g_i : \N \to \R_+ , i \in I.$ All functions $ \phi_i $ and $ g_i$ are measurable functions. We assume that $\phi_i$ is increasing and uniformly Lipschitz continuous, i.e.\ there exists a positive constant $ \gamma $ 
such that for all $ s, s' \in \R ,$ 
\begin{equation}\label{eq:Lip}
| \phi_i (s ) - \phi_i  ( s'  ) | \le \gamma  |s - s' |   .
\end{equation}

We start by considering a model in discrete time which is partly inspired by Cessac (2011)\nocite{cessac_discrete_time} who proposes a finite dimensional system. We consider a stochastic chain $(X_t  )_{ t \in \Z }$ taking values in $ \{ 0, 1 \}^I ,$ where $I$ is the countable set of neurons, defined on a suitable probability space $ ( \Omega , { \cal A} , P ) .$ For each neuron $i$ at each time $t \in \Z,$  $ X_t (i) = 1 $ reports if neuron $i$ has a spike at that time $t.$ Otherwise we put $X_t(i) = 0 .$ The global configuration of neurons at time $t$ is denoted $X_t = (X_t (i) , i \in I ) .$ For each neuron $i \in I$ and each time $t \in \Z $ let
\begin{equation}
L_t^i = \sup \{ s < t  : X_s (i) = 1  \} 
\end{equation} 
be the last spike time of neuron $i$ strictly before time $t.$ At each time $t ,$ conditionally on the whole past, sites update independently. This means that for any finite subset $J \subset I ,$ $a_i \in \{ 0, 1 \} , i \in J, $ if we introduce the filtration 
$$ \F_t = \s ( X_s ,  s \in \Z, s \le t   ) , t \in \Z ,$$ 
then we have
\begin{equation}\label{eq:dyn}
P ( X_t (i ) = a_i , i \in J | \F_{t-1} ) = \prod_{i \in J} P ( X_t (i ) = a_i | \F_{t-1}),
\end{equation}
where
\begin{equation}\label{eq:transition2}
 P( X_t (i ) = 1 | { \cal F}_{t- 1}  ) = \phi_i  \left( \sum_j  W_{ j \to i} \sum_{ s = L_t^i}^{ t- 1} g_j (t-s) X_s (j)    \right) .
\end{equation}
Here $\phi_i $ is the spiking rate function of neuron $i, $ introduced above, and $g_j $ the leak function associated to the spiking events of the $j-$th neuron. Observe that, since $\phi_i $ is increasing, the contribution of components $j$ is either excitatory or inhibitory, depending on the sign of $W_{j \to i} .$

\section{Infinite systems of interacting processes with memory of variable length in continuous time}\label{sec:continuous time}
In continuous time, the activity of each neuron $i \in I$ is described by a counting process $ Z^i $ recording for any $ - \infty < s < t < \infty $ the number $Z^i (]s, t ]) $ of spikes of neuron $i$ during the interval $ ]s, t ]. $ The sequence of counting processes $ ( Z^i , i \in I) $ is characterized by its intensity process $ (\lambda_t^i , i \in I) $ defined through the relation 
$$ P ( Z^t \mbox{ has a jump in [t , t + dt ]} | \F_t ) = \lambda_t^i dt , i \in I,$$
where $ \F_t = \sigma (  Z^i ( ] s, u ] ) , s \le u \le t , i \in I) $
and where 
\begin{equation}\label{eq:intensity1}
 \lambda^i_t  =  M_i \, \phi_i \left( \sum_{j \in I}  W_{j \to i }  \int_{[L_t^i , t [} g_j ( t-s) d Z^j_s  \right)  .
\end{equation}
Here, $M_i, i \in I, $ is a collection of positive numbers giving the maximal intensity of spiking per neuron (recall that $\phi_i $ takes values in $[0, 1 ]$), and $L_t^i = \sup \{ s < t : Z^i ( [s] > 0 \} .$

This form of  an intensity process is close to the typical form of the intensity of a multivariate nonlinear Hawkes process as it has been considered since Hawkes (1971)\nocite{hawkes71}. We refer the interested reader to Br\'emaud and Massouli\'e (1996)\nocite{BremaudMassoulie96} for an extensive and comprehensible study of stability properties of nonlinear Hawkes process, and to Hansen, Reynaud-Bouret and Rivoirard \nocite{hrbr} for the use of Hawkes processes as models of spike trains in neuroscience. See also Delattre, Fournier and Hoffmann (2014)\nocite{dfh} for a study of infinite systems of nonlinear Hawkes processes. Our form of the intensity (\ref{eq:intensity1})  differs from the classical Hawkes setting by its {\it variable memory structure} introduced through the term $ L_t^i .$ Hence the spiking intensity of a neuron only depends on its history up to its last spike time which is a biologically very plausible assumption on the memory structure of the process. Therefore, our model can be seen as a nonlinear multivariate Hawkes process where the number of components is infinite with a variable memory structure.

\section{Associated Markov interacting particle system}\label{sec:Markov}
The choice of a leak function $g_j \equiv 1$ in \eqref{eq:intensity1} gives rise to an intensity process which is Markov. In this case, write 
\begin{equation}
 U_i ( t) = \sum_{j \in I}  W_{j \to i }  \int_{[L_t^i , t [} d Z^j_s .
\end{equation}
We can interpret $U_i ( t)$ as value of the membrane potential of neuron $i$ at time $t.$ 
Then it is straightforward to see that $ (U_i (t) )_{i \in I} $ is a Markov process taking values in $\R^I, $
whose generator is given for any smooth test function $ f : \R^I \to \R $ by
\begin{equation}\label{eq:generator0}
L f (x ) = \sum_{ i \in I } M_i \phi_i (x_i) \left[f (x + \Delta_i ( x)  ) -f (x) \right]
,
\end{equation}
where
\begin{equation}
(\Delta_i (x))_j =    \left\{
\begin{array}{ll}
W_{i \to j } & j \neq i \\
- x_i & j = i
\end{array}
\right\} .
\end{equation}
In such a system of interacting processes $U_i  (t), $ each neuron is represented by the height $x_i$ of its membrane potential. It spikes at a rate depending on this height. When spiking, it goes back to the value $0$ which can be interpreted as resting potential. At the same time, neurons influenced by $i, $ i.e.\ the postsynaptic neurons, receive an additional amount of potential $ W_{ i \to j }, $ independently of the former value $x_i$  of the membrane potential of the spiking neuron. In particular, there is no conservation of mass (i.e.\ of potential), since the spiking neuron does not re-distribute its own potential (this is for instance a main difference with the Potlatch process or with sandpile processes).

Notice that from a mathematical point of view the existence of such a process in infinite dimension, i.e.\ in the case when $ I$ is infinite, is not evident, since the interactions might come down from infinity. We do not go into the details, but for a general discussion of existence issues in infinite dimension, we invite the interested reader to consult for example Chapter 1 of Liggett (1985). \nocite{Liggett1985}

Sometimes, we will  concentrate on the finite case and take $I  = \{ 1, \ldots , N\} ,$ for some fixed $ N > 0.$ In this case, we might add to the above dynamics \eqref{eq:generator0} two terms. The first one is a leak term modeling the fact that throughout its evolution, the membrane potential looses potential due to leakage channels. The second is a drift term modeling the effects of gap-junctions to the system. Whereas the leakage channels tend to push the membrane potential of each neuron towards zero, the gap junctions, on the contrary, tend to push the whole system towards its average membrane potential value. We are thus led to consider a continuous time Markov process $ U(t) = (U_1 ( t) , \ldots , U_N ( t) ) $ taking values in $\R^N, $ whose infinitesimal generator is given for any smooth test function $ f: \R^N \to \R $ by
\begin{equation}\label{eq:generator1}
L f (x ) = \sum_{ i =1 }^N  M_i \phi_i (x_i) \left[f (x + \Delta_i ( x)  ) - f (x) \right] - \lambda \sum_{i=1}^N \frac{\partial f}{\partial x_i } (x) [ x_i - \bar  x ] - \alpha \sum_{i=1}^N \frac{\partial f}{\partial x_i} (x) x_i 
,
\end{equation}
where $\lambda, \alpha \geq 0 $ are positive parameters and where $ \bar x = \frac1N \sum_{i=1}^N x_i .$ Here, $\lambda $ models the strength of electrical synapses, and $\alpha $ the leakage effect. 

\section{Existence results and perfect simulation}\label{sec:results1}
It is natural to ask if there exists at least (and at most) one stationary process which is consistent with the dynamics defined through \eqref{eq:dyn} and \eqref{eq:transition2} in discrete time or through \eqref{eq:intensity1} in continuous time. 
The answer to this question is intimately related to the structure of interactions given by the synaptic weights. These interactions  can be represented as a directed weighted graph where the directed link $ i \to j$ is present if and only if $W_{ i \to j } \neq  0,$ and where each directed link is weighted by $ W_{ i \to j } .$ For each neuron $i,$ we introduce
$$ {\cal V}_{\cdot \to i} = \{ j \in I, j \neq i  : W_{j \to i } \neq 0 \} ,$$
the set of all neurons that have a direct influence on neuron $i.$ Notice that in our model,  ${\cal V}_{\cdot \to i}  $ can be both finite or infinite. We fix a growing sequence $( V_i (k ) )_{ k \geq - 1 } $ of subsets of $ I $ such that $V_i ( - 1 ) = \emptyset ,$ $ V_i ( 0 ) =  \{ i \}  ,$ $ V_i (k ) \subset  V_i ( k+1) ,$ $  V_i (k ) \neq V_i ( k+1)$ if $ V_i (k ) \neq {\cal V}_{\cdot \to i}   \cup \{ i \}$ and $\bigcup_k V_i (k) ={\cal V}_{\cdot \to i}  \cup \{ i \} .$

We now state our existence and uniqueness result. We formulate it for discrete time systems incorporating spontaneous spike times, see Condition (\ref{eq:delta}) below. These spontaneous spikes can be interpreted as external stimulus or, alternatively, as autonomous activity of the brain. In order to state our result, let us introduce, for all 
$ s < t \in \Z, $ the process $ X_s^t (i) = ( X_s (i), X_{ s+1} ( i ) , \ldots , X_t (i) ) ,$ which is the trajectory of $X(i)$ between times $s$ and $t.$

\begin{theo}\label{theo:3}[Theorem 1 of Galves and L\"ocherbach (2013)]\\
Grant conditions (\ref{eq:summable}) and (\ref{eq:Lip}). Assume that the functions $\phi_i  $ and $g_j$ satisfy moreover the following assumptions:

i) There exists $ \delta > 0 $
such that  for all $ i \in I, s \in \R   ,$ 
\begin{equation}\label{eq:delta}
\phi_i  ( s ) \geq \delta .
\end{equation}
ii)
We have that
\begin{equation}\label{eq:G}
  G (1) + \sum_{ n = 2 }^\infty ( 1 - \delta)^{ n - 2} n^2  G ( n) < \infty ,
\end{equation}
where $G  ( n) = \sup_i \sum_{ m = 1}^n  g_i (m) $ and where $ \delta $ is as in condition 1.

iii)
We have fast decay of the synaptic weights, i.e.
\begin{equation}\label{eq:veryfast}
\sup_i \, \sum_{ k \geq 1 } | V_i (k ) | \left( \sum_{ j \notin V_i (k - 1 ) } |W_{j \to i }| \right) < \infty . 
\end{equation}

Then the following assertions hold true. \\
1) There exists a critical parameter $\delta_* \in ] 0, 1 [ $ such that for any $ \delta > 
 \delta_* , $ there exists a unique probability measure $P  $ under which $(X_t)_{t \in \Z}$ satisfies (\ref{eq:dyn}) and (\ref{eq:transition2}). \\
2) There exists a non increasing function $\ell : \N \to \R_+ ,$
such that for any $ 0 < s < t \in \N $ the following holds. For all $i \in I, $ for all bounded measurable functions $f : \{ 0, 1 \}^{   [ s, t ] } \to \R_+ ,$
\begin{equation}\label{eq:lossofmemory}
 \big| E [ f ( X_s^t (i) )  | {\cal F}_0  ] - E [ f ( X_s^t (i))] \big| \le   \, ( t-s +1 ) \, \| f \|_\infty \, \ell (s)    .
\end{equation}
Moreover, $\ell ( n)  \le C \frac{1}{n- 1} $ for some fixed constant $C. $ 
\end{theo}

\begin{ex}
Take $ I = \Z^d ,$ $g_j ( s) = 1$ for all $j, s,$ and 
$$ W _{ i \to j } = \frac{ 1 }{ \|j- i\|_1^{2 d + \alpha} }$$ for some fixed $\alpha > 1,$ where $\| \cdot \|_1$ is the $L^1-$norm on $\Z^d .$  In this case, if we choose $ V_i (k ) = \{ j \in \Z^d = \| j - i \|_1 \le k \}, $ we have $ | V_i (k ) | = (2k + 1)^d , $ and it is easy to see that condition (\ref{eq:veryfast}) is satisfied.
\end{ex}

The proof of Theorem \ref{theo:3} implies the existence of a perfect simulation algorithm of the stochastic chain $(X_t)_{t \in \Z} .$ By a perfect simulation algorithm we mean a simulation which samples in a finite space-time window precisely from the stationary law $P .$ We refer the interested reader to Galves and L\"ocherbach (2013) \nocite{antonioeva13}. In continuous time, the existence of a unique stationary version of a process $(Z^i )_{i \in I}$ having intensity \eqref{eq:intensity1} can be shown by following similar same ideas. Details can be found in Hodara and L\"ocherbach (2014)\nocite{PE}. In particular, here again, we obtain a perfect simulation algorithm for the stationary law.

\section{The interaction graph and de-correlation of neighboring interspike intervals }\label{sec:results2}
Throughout this chapter, we work within the discrete time model of Section 2.

One central question in theoretical neuroscience is the distribution of consecutive interspike intervals (ISI) and in particular their dependence or independence. In order to answer to this question, we have to specify our choice of an interaction graph. This is related to the second central question in theoretical neuroscience: what kind of graph should be considered? Beggs and Plenz (2003)\nocite{BeggsandPlenz} argue that networks of living neurons should behave in a slightly supercritical state. Therefore we consider a slightly supercritical directed Erd\"os-R\'enyi random graph.

More precisely, for a large but finite system of $N$ neurons, let $ W_{i \to j} , i \neq j  , 1 \le i, j \le N,$  be random synaptic weights. The sequence $ W_{i \to j} , i \neq j ,$ is a sequence of i.i.d.\ Bernoulli random variables defined on some probability space $ ( \tilde \Omega , \tilde {\cal A} , \tilde P ) $ with parameter $p = p_N  ,$ 
i.e. 
$$ \tilde P ( W_{i \to j}  = 1) = 1  - \tilde P(  W_{i \to j}  = 0 ) = p_N ,$$
where 
\begin{equation}\label{eq:criticalregime}
 p_N = \lambda / N \mbox{ and } \lambda = 1 + \vartheta / N \mbox{ for some } 0 < \vartheta < \infty .
\end{equation}
We put $ W_{j \to j } \equiv 0 $ for all $ j .$ Conditionally on the choice of the connectivities $ W = ( W_{ i \to j }, i \neq j ) ,$ the dynamics of the chain are then given by
$$ P^W ( X_t (i) = 1 | {\cal F}_{t-1} ) =   \phi_i ( \sum_{ j } W_{j \to i } \sum_{ s= L_t^i }^{t- 1} g_j ( t-s) X_s (j)  )$$
as before. We denote $P^W $ the conditional law of the process, conditioned on the choice of $W .$

Fix a neuron $i$ and consider its associated sequence of successive spike times 
\begin{equation}\label{eq:spikes}
 \ldots < S^i_{- n } < \ldots < S^i _0 \le 0 < S^i_1 < S^i_2 < \ldots < S^i_n < \ldots ,
\end{equation}
where 
$$ S_1^i = \inf \{ t \geq 1 : X_t(i) = 1 \} , \ldots , S_n^i = \inf \{ t > S_{n-1}^i : X_t (i ) = 1 \} , n \geq 2 , $$
and
$$ S_0^i = \sup \{ t \le 0 : X_t (i) = 1 \} , \ldots , S^i_{ -n} = \sup \{ t < S^i_{ - n + 1 } : X_t (i) = 1 \} , n \geq 1 .$$

Let us fix $W.$ We are interested in the covariance between successive inter-spike intervals
$$ Cov^W (  S^i_{k+1}- S^i_k , S^i_k - S^i_{k-1} ) = E^W [( S^i_{k+1}- S^i_{k} ) ( S^i_k - S^i_{k-1})] - E^W ( S^i_{k+1} - S^i_k ) E^W ( S^i_k - S^i_{k-1} )   ,$$
for any $ k \neq 0,   1 .$
Being in stationary regime, the above covariance does not depend on the particular choice of $k .$ The next theorem shows that neighboring inter-spike intervals are asymptotically uncorrelated as the number of neurons $N$ tends to infinity.  

\begin{theo}\label{theo:4}[Theorem 3 of Galves and L\"ocherbach 2013]
Assume that (\ref{eq:Lip}), (\ref{eq:delta}) and (\ref{eq:G}) are satisfied. Then there exists a measurable subset $ A \in \tilde {\cal A} ,$ such that on $A,$ 
$$ |Cov^W (  S^i_3- S^i_2 , S^i_2 - S^i_1 )| \le \frac{3}{\delta^2} N (1 - \delta)^{\sqrt{N} }  , $$
where $\delta $ is the lower bound appearing in Condition (\ref{eq:delta}). 
Moreover,  
$$ \tilde P ( A^c  ) \le e^{2\vartheta} N^{ - 1/2}  .$$

\end{theo}

For large $N,$ if the graph of synaptic weights belongs to the ``good'' set $A,$ the above result is compatible with the discussion in Gerstner and Kistler (2002) arguing that two consecutive interspike intervals can be considered as independent.

\section{Longtime behavior for the associated Markov interacting particle system}\label{sec:longtime}
We briefly discuss the longtime behavior of the Markov interacting particle system having generator \eqref{eq:generator1}. Notice that such a process is a {\it piecewise deterministic Markov process} (PDMP). We concentrate on the case when all synapses are excitatory, i.e.\ $W_{ i \to j } \geq 0 $ for all $i , j .$ 
Moreover we consider a homogenous population where all neurons have the same spiking behavior, i.e. $ M_i \phi_i \equiv \phi $ for all $i.$ We do not suppose $\phi$ to be bounded nor to be globally Lipschitz continuous any more. All we have to assume is 
\begin{ass}\label{ass:1}
$\phi : \R_+ \to \R_+$ is non-decreasing, $\phi(0) = 0$, $\phi (x)>0$ for all $x>0$, there exists $ r > 0$ such that $ \int_0^{2r} \phi (x)/x dx < \infty ,$ $\lim_\infty \phi  = \infty$ 
and $\phi \in C^1(\mathbb R_+)$.
\end{ass}
In this case, the existence of the process is deduced by a simple coupling argument going back to Fournier and L\"ocherbach (2014) \nocite{evanicolas} proving that the total number of jumps during any finite time interval is finite almost surely. Moreover, interestingly enough, Duarte and Ost (2014)\nocite{bresiliens} show that, if the parameter $r $ of Assumption \ref{ass:1} satisfies
$$ r > \max_i \sum_j W_{j \to i } , $$
and if $ \alpha > 0,$ i.e.\ there is some leakage phenomenon, then 
$$ P ( \exists T \mbox{ such that no spikes occur in } [T, \infty [ ) = 1 .$$ 
(Theorem 2.3 of Duarte and Ost (2014), compare also to Theorem 1 of Robert and Touboul (2014)\nocite{Touboul-Robert}). In particular, in this case, the process goes extinct almost surely. However, if $\alpha = 0$ and there is no leak effect, then it is straightforward to show that the process does not go extinct, but will converge to a non trivial invariant measure. Even more, in this case the process is recurrent in the sense of Harris on $\R_+^N \setminus \{0 \} , $ see Theorem 2.4 of Duarte and Ost (2014). In particular, the trivial invariant measure $\delta_0$ is non attractive.

\section{Mean field limits}
Suppose we observe a large homogeneous population of $N$ neurons in continuous time evolving according to \eqref{eq:generator1}. Then we can assume that we are in an idealized situation where all neurons have identical properties, leading to a mean field description. The mean field assumption appears through the assumption that $ W_{i \to j } = \frac1N $ for all $ i \neq j .$ Moreover, we suppose from now on, that there is no leakage effect, i.e.\ $\alpha = 0 .$ 
In order to keep track of the size of the system, we denote the process by 
$$U^N (t) = (U^N_1 (t), \ldots , U^N_N ( t) )   ,  \, t \geq 0 , $$
and identify the state of the system at time $t$ with its empirical measure
\begin{equation}
\label{eq:emp}
\mu^N_{ t}  = \frac 1N \sum_{i=1}^N \delta_{U^N_i(t)} .
\end{equation}

In Theorem 2 of De Masi et al.\ (2015)\nocite{AAEE} it has been shown that, in the limit as $N\to \infty,$
this membrane potential distribution becomes deterministic and it is described by a density $\rho_t(r),$ where for any  interval $I \subset \mathbb R _+$, $\int_I\rho_t(r)dr$ is the limit fraction of neurons whose membrane potentials are in $I$
at time $t$.  The limit density  $\rho_t(r)$ is proved to obey a non linear PDE which is a conservation law of hyperbolic type
\begin{equation}
 \label{eq:217.3}
\frac{ \partial }{ \partial t}\rho_t + \frac{ \partial }{ \partial x}(V\rho_t) = - \phi \rho_t,\quad x>0,t>0 ,
\end{equation}
where 
\begin{equation}
 \label{eq:217.2}
 V(x,\rho_t):=  -\lambda  (x - \bar\rho_t)+ p_t 
  \end{equation}
is the velocity field, where the first term describes the attraction to the average membrane potential of the system, due to the gap junction effect, the second one the drift produced by the other neurons spiking. Here, the limit total firing rate per unit time $p_t$  and  the limit average membrane potential $\bar \rho_t$ are
\begin{equation}
\label{eq:217.1}
 p_t = \int_0^\infty \phi(x) \rho_t (x) dx, \;\;\;\; \bar \rho_t = \int_0^\infty x \rho_t (x) dx .
\end{equation}
In \eqref{eq:217.3}, the expression $- \phi (x) \rho_t(x)$ is a  loss of mass term due to spiking. Finally, since \eqref{eq:217.3} is only defined for $ x , t > 0, $ we have to complete the PDE with boundary conditions which read as follows.
\begin{equation}
 \label{eq:217.3.1}
\rho_0(x)= u_0(x),\quad  \rho_t(0)= \frac{p_t}{V(0,\rho_t)} = \frac{p_t}{p_t+ \lambda \bar\rho_t} ,
   \end{equation}
where $u_0$ is some initial density of neurons at time $t=0.$ 

It can be shown that under suitable assumptions on the initial distribution of neurons at time $0,$ there exists a unique weak solution $\rho_t (x) $ of  \eqref{eq:217.3}--\eqref{eq:217.3.1}.
This is e.g.\ the case if the distribution of neurons at time $0$ is of compact support. For further details, we refer the reader to Theorem 4 of Fournier and L\"ocherbach (2014). 

\begin{theo}[Theorem 2 of De Masi et al.\ (2015)]
\label{theo:main}
Grant Assumption \ref{ass:1} and suppose that $U_i^N ( 0), 1 \le i \le N$ are i.i.d. random variables having smooth density $u_0 ( x) .$ Let $\rho_t (x) $ be the unique weak solution of  \eqref{eq:217.3}--\eqref{eq:217.3.1}. Then for any fixed $ T > 0 , $
\begin{equation}
{\cal L} (\mu_{ U^{N}_{ [0, T ] }} ) \stackrel{w}{\to} {\cal P}_{[0, T ] }
\end{equation}
(weak convergence in $ D ( [0, T ]  , {\cal S}') $) as $N \to \infty ,$ where $ {\cal P}_{ [0, T ] } $ is the law on $  D ( [0, T ] , {\cal S}') $ supported by the distribution valued trajectory $ \omega_t $ given by
$$ \omega_t ( \phi ) = \int_0^\infty  \phi (x) \rho_t (x) dx , \quad  t \in [0, T ] ,$$
for all $\phi \in {\cal S}.$
\end{theo}

\begin{rem}
The equivalence between the ``chaoticity'' of the system and a weak law of large numbers for the empirical measures, as proven in Theorem \ref{theo:main}, is well-known (see for instance Sznitman (1999)).This means that in the large population limit, the neurons converge in law to independent and identically distributed copies of the same limit law. This property is usually called ``propagation of chaos'' in the literature.
\end{rem}

In case $\lambda = 0  $ and $u_0 ( 0) = 1,$  \eqref{eq:217.3} reads as follows.
$$ \left\{
\begin{array}{lcll}
\partial_t \rho_t(x)    &=&- p_t  \partial_x \rho_t(x)    - \phi(x)\rho_t (x) , &  x > 0 ,
\\
 \rho_t(0 ) &= & 1  \;&  \mbox{for all $t \geq 0$}  .
\end{array}
\right. $$
This equation is different from the so-called ``population density equations'' which are obtained for integrate-and-fire neurons as considered e.g.\ in Chapter 6.2.1 of Gerstner and Kistler (2002), see in particular their formula (6.14). As in integrate-and-fire models, also in our model spiking neurons are reset to a reversal potential (which equals $0$); but spiking does not create Dirac-masses at the reset value. This is due to the Poissonian mechanism giving rise to spiking in our model. The loss of mass at time $t$ due to spiking of neurons having potential height $x$ is therefore described by the term $ - \phi (x) \rho_t (x) .$

At the same time, spiking induces a deterministic drift $ p_t dt $ for those neurons that are not spiking. Finally, 
conservation of total mass implies that the initial density of neurons at the border $x= 0 $ is of height $1.$ This initial condition is different from the usual initial condition obtained in integrate-and-fire models.

\begin{rem}
The mean field approach intending to replace individual behavior in large homogeneous systems of interacting neurons by the mean behavior of the neuronal population has a long tradition in the frame of neural networks, see e.g.\ Chapter 6 of Gerstner and Kistler (2002)\nocite{Gerstner} or Faugeras et al.\ (2009)\nocite{Faugerasetal2009} and the references therein. Most of the models used in the literature are either based on rate models where randomness comes in through random synaptic weights (see e.g.\ Cessac et al.\ (1994)\nocite{Cessac1994} or Moynot and Samuelides (2002)\nocite{Moynot02}); or they are based on populations of integrate and fire neurons which are diffusion models in either finite or infinite dimension, see for instance Delarue et al.\ (2012)\nocite{delarue} or Touboul (2014)\nocite{Touboul2014}.  The model we consider is reminiscent of integrate-and-fire models but firing does not occur when reaching a fixed threshold, and the membrane potential is not described by a diffusion process. The only noise which comes in is the Poissonian noise given by the spiking features, compare also to Robert and Touboul (2014).\nocite{Touboul-Robert} 
\end{rem}

\section{Further results}\label{sec:further results}
The limit density $\rho_t (x) $ which is solution of \eqref{eq:217.3}--\eqref{eq:217.3.1} can be interpreted as density of a typical single neuron $U(t) ,$ evolving within an infinite system of neurons according to \eqref{eq:generator0}. Its dynamics can be described as follows. Let $U (0)$
be a $u_0$-distributed random variable, 
independent of a Poisson measure $N(ds,dz)$ on 
$\R_+ \times \R_+ $ having intensity measure $ ds dz$. Then 
\begin{multline}\label{eq:dynlimit}
U (t) = U(0) - \lambda \int_0^t (U(s)  - E[ U(s)])ds \\
- \int_0^t\int_0^\infty U({s- }) 1_{ \{ z \le  \phi  ( U({s-})) \}} N  (ds, dz) + \int_0^t E[\phi (U(s))]ds.
\end{multline} 
Notice that the above dynamics is the dynamics of a nonlinear Markov process (in particular, non homogenous in time), since the law of the process itself -- representing the state of the other neurons within the infinitely large system --  is involved in its dynamics.

As in the case of finite systems of neurons, also the limit process possesses exactly two invariant measures
supported in $\R_+$. The first one is $\delta_0 $. The second one is of the form $g(dx)=g(x)dx$,
with $g:[0,\infty) \mapsto [0,\infty)$ defined by
$$ g (x) = \frac{p}{p + \lambda m - \lambda x } 
\exp \Big( - \int_0^x \frac{\phi (y) }{p + \lambda ( m - y ) } dy \Big) 1_{\{ 0 \le x < m + p/ \lambda \} },$$
where $p>0$ and $m>0$ are uniquely determined by the constraints
$\int_0^\infty g(dx)= 1$, $\int_0^\infty x g(dx)= m$.
Furthermore, we have $\int_0^\infty \phi (x) g(dx) = p$ and  $m +p/\lambda>1$.
Note that for $\lambda =0 ,$ this reads as  
\begin{equation}\label{eq:invariant}
 g(x) = \exp \Big( - \frac{1}{p}\int_0^x \phi (y) dy \Big) .
\end{equation}

Contrary to the case of a finite system of neurons, it is surprisingly difficult to show that the limit system does not go extinct, i.e.\ $\rho_t (x) dx $ does not tend to $\delta_0, $ weakly - at least in the case of presence of gap junctions $ \lambda > 0 .$ The whole picture is only known in the case $ \lambda = 0 .$ 

\begin{prop}[Prop. 9 of Fournier and L\"ocherbach (2014)]
Grant Assumption \ref{ass:1} and suppose moreover that $\phi  \in C^2(\R_+)$ is convex increasing and 
$\sup_{x\geq 1} [\phi'(x)/\phi(x)+\phi''(x)/\phi'(x)]<\infty$. Let $\lambda=0$ and suppose that $U(0) \sim  u_0 \in C^1_b([0,\infty))$
where $u_0$ satisfies
$u_0(0)=1$, $\int_0^\infty \phi^2 (x) u_0 (x) dx 
< \infty $ and $\int_0^\infty |u_0'(x)| dx < \infty$.  
Denote by $\rho(t)$ the law of $U(t)$ and write $g (dx) = g(x) dx  $ for the invariant probability measure
defined in \eqref{eq:invariant}. Then we have $\lim_{t\to\infty} \|\rho( t) -g\|_{TV}=0,$ where 
$ \|  \cdot \|_{TV} $ denotes the total variation distance. In particular, the process does not go extinct, almost surely. 
\end{prop}

The case $\lambda \neq 0$ is more subtle, and we only know that the process does not go extinct, under minimal conditions on the spiking rate function, cf. Proposition 11 of Fournier and L\"ocherbach (2014). 

\section{Questions and challenges}\label{sec:questions}
In this section we raise several natural questions in the context of the models considered in this paper.

{\it To which extend is a mean field description as adopted in Sections 8 and 9 above relevant from a neurobiological point of view?} The mean field approach intending to replace individual behavior in large homogeneous systems of interacting neurons by the mean behavior of the neuronal population has a long tradition in the frame of neural networks. 
Bressloff (2009)\nocite{bressloff} argues that considering 
homogeneous populations ``is motivated by the observation that neurons in cortex with similar response properties tend to be arranged in vertical columns. A classical example is the columnar-like arrangement of neurons in primary visual cortex that prefer stimuli of similar orientation''. Therefore it is reasonable to consider that such systems of neurons are governed by interactions of mean-field type.  

Moreover, Bojak et al.\ (2010)\nocite{bojak} claim ``that a mean field model of brain activity can simultaneously predict EEG and fMRI BOLD ...''. For a recent review paper we refer the reader to Pinotsis et al.\ (2014).\nocite{pinotsis}

{\it Description of the system and propagation of chaos.}
EEG as well as fMRI data describe the collective behavior of huge subpopulations of neurons. This makes it reasonable to consider a space-time rescaling of 
the ``microscopic system'' reminiscent of what is usually done for interacting particle systems under the name of ``hydrodynamical limits'', see De Masi and 
Presutti (1991)\nocite{anna-errico91} and Kipnis and Landim (1999)\nocite{kipnis-landim}. The difficulty for neurobiological models is that contrarily to the case of thermodynamical systems considered in statistical physics, we have no macroscopic qualitative results available. In a nutshell, as far as we know, in neurobiology there is presently nothing that plays the role that the Fourier law plays in thermodynamics. Therefore, the first problem is to understand what kind of limiting behavior should be obtained when rescaling a stochastic model describing a system of spiking neurons. 

Chaos propagation is an issue which is directly associated to the above discussion. The concept of propagation of chaos has been introduced by Mark Kac in his seminal paper of 1956, see Kac (1956).\nocite{Kac}. His goal was to show that within a system of a huge number of interacting components, in a suitable limit, any fixed number of components behave as independent stochastic processes having the same distribution (for more details, we refer the reader to the classical reference Sznitman (1991)\nocite{sznitman}). 

The recent literature in neuromathematics presents many results concerning the propagation of chaos in stochastic models of neuronal networks. However, it is far from being clear what is the neurobiological meaning of these results. Moreover, it is difficult to find neurobiological experimental results clearly related with chaos propagation. At the same time, in a large majority of the mathematical papers, including ours, there is no discussion about the neurobiological relevance of this issue. 

Exceptions are recent papers and lectures of Olivier Faugeras and members of his team. For instance, in  
Baladron et al.\ (2012) \nocite{baladronetal} the authors write the following.``We prove a propagation of chaos property which shows that in the mean-field limit, the neurons become independent, in agreement with some recent experimental work [13] and with the idea that the brain processes information in a somewhat optimal way.'' In the above, [13] refers to Ecker et al.\ (2010) which present experimental evidence concerning the de-correlation of neuronal firing in the visual cortex. Here, the authors argue that ``the de-correlated state of the neocortex [...] offers substantial advantages for information processing.''  \nocite{Ecker}

We believe that a more systematical discussion of the relation between mathematical results on chaos propagation and qualitative experimental results of neurobiology should be done by the neuromathematical community.

{\it What about inhibitions?} Inhibitions are considered in Sections 2, 3 and 5, but the theoretical results proved there do not take advantage of the
balance between excitatory and inhibitory neurons. As far as we know, no rigorous neuromathematical result relies on this balance. This is clearly a
step which must be achieved in a near future. 

On the other hand, in many articles of the neurobiological literature it is very often suggested that inhibition should play a crucial role to explain 
many qualitative aspects found in the data. For instance, Benayoun et al.\ (2010) \nocite{benayoun}suggest that the balance between inhibition and excitation is an important ingredient in the explanation of avalanche phenomena. However, it is quite simple to conceive mathematical toy models without inhibition for systems of spiking neurons in which avalanches are produced. Another example is the belief which seems to be widespread in the neuroscientific community
concerning the role played by inhibition in phenomena like chaos propagation (cf. for instance, van Vreeswijk and Sompolinsky (1996) and 
Huntsman et al.\ (1999)). \nocite{Vreeswijk} \nocite{Huntsman} However, our Theorem \ref{theo:main} concerning propagation of chaos does not
rely at all on the presence of inhibitory synapses. We believe that it is important to better understand the importance of inhibition in the qualitative
behavior of stochastic models like those presented in this paper.

{\it What about statistical model selection?} We have presented in this paper what we believe to be a good class of candidate models
describing spiking neuronal behavior. Therefore, if we were able to do statistical model selection for this class of models we would be able to clarify 
issues like the way different neurons and regions of the cortex interact. In mathematical terms, this amounts of {\it estimating} the underlying interaction graph. There are two obvious difficulties. 

First of all, our data concerns only an extremely tiny part of the cortex (between $10^2$ to $10^3$ neurons).
What does this very small view tell us about the entire region? In former papers, one of the authors show that for models coming from statistical physics like the Ising model, under a Dobrushin type condition, it is possible to make inference of the entire system by just observing a small part of it. We refer the interested reader to Galves, Orlandi and Takahashi (2015)\nocite{enzaantonioetal}. Is this type of result still valid when we look at a system like the brain? 

The second problem is the computational complexity of the selection procedure among all possible interaction graphs supporting models like ours. 
In the case of chains with memory of variable length, the famous Context algorithm introduced by Rissanen (1983)\nocite{rissanen} and more recent versions of the BIC approach to the same problem by Cszisar and Talata (2006) \nocite{csiszar2006}as well as its application to random Markov fields of variable neighborhoods in L\"ocherbach and Orlandi (2011)\nocite{evaenza}, this selection problem can be reduced to a linear complexity problem. Is a solution like this available for the class of models considered here?

\section{Discussion}\label{sec:discussion}
We close with a discussion of the particular aspects of the models we propose in this paper. We start by discussing the `` variable length memory dependence'' on the past which is incorporated in our models via \eqref{eq:transition2}  or \eqref{eq:intensity1}. From a theoretical point of view, Cessac (2011)\nocite{cessac_discrete_time} suggested the same kind of dependence from the past in a discrete time model. In the framework of leaky integrate and fire models, he considers a system with a finite number of membrane potential processes. The
image of this process in which only the spike times are recorded is a stochastic chain of infinite order where each neuron has to look back into the past until its last spike time. Cessac's process is a finite dimensional version of the model considered in Section 2. 

A second important feature of the class of processes introduced in our paper is the fact that we are able to deal with infinite systems of neurons in a rigorous mathematical way. 
Finite systems of point processes in discrete or continuous time aiming to describe biological neural systems have a long history whose starting points are probably Hawkes (1971)\nocite{hawkes71} from a probabilistic point of view and  Brillinger (1988)\nocite{Brillinger1988} from a statistical point of view, see also the interesting paper by Krumin et al. (2010)\nocite{krumin} for a review of the statistical aspects. For non-linear Hawkes processes, but in the frame of a finite number of components, Br\'emaud and Massouli\'e (1994)\nocite{BremaudMassoulie96} address the problem of existence, uniqueness and stability. M{\o}ller and coauthors propose a perfect simulation algorithm in the linear case, see M{\o}ller and Rasmussen (2005)\nocite{MollerRasmussen}. 
In spite of the great attention that Hawkes processes received recently, especially in association with modeling problems in finance and biology, all the studies are reduced to the case of systems with a finite number of components. Only recently, Delattre, Fournier and Hoffmann (2014)\nocite{dfh} studied an infinite system of nonlinear Hawkes processes but did not address the perfect simulation issue. Notice also that in none of the above articles, a variable length memory dependence on the past is incorporated although this kind of dependence is completely natural in the frame of neural nets.  

\section*{Acknowledgments}
We thank Pierre Hodara and Daniel Takahashi for careful reading of this manuscript and valuable comments. 

This article was produced as part of the activities of FAPESP Research,
Dissemination and Innovation Center for Neuromathematics (grant
2013/07699-0, S.\ Paulo Research Foundation). This work is part of USP project ``Mathematics, computation, language
and the brain" and CNPq project ``Stochastic modeling of the brain activity"
(grant 480108/2012-9). AG is partially supported by a CNPq fellowship
(grant 309501/2011-3). 

\bibliography{biblio}

\begin{thebibliography}{10}

\bibitem{baladronetal}
J.~Baladron, D.~Fasoli, O.~Faugeras, and J.~Touboul.
\newblock {Mean-field description and propagation of chaos in networks of
  Hodgkin-Huxley and Fitzhugh-Nagumo neurons.}
\newblock 2012.

\bibitem{BeggsandPlenz}
J.M. Beggs and D.~Plenz.
\newblock Neuronal avalanches in neocortical circuits.
\newblock {\em J. Neurosc}, 23:11167--11177, 2003.

\bibitem{benayoun}
M.~Benayoun, J.D. Cowan, W.~van Drongelen, and E.~Wallace.
\newblock {Avalanches in a Stochastic Model of Spiking Neurons.}
\newblock {\em PLOS Comput Biol.}, 6(7), 2010.

\bibitem{bojak}
I.~Bojak, T.F. Oostendorp, A.T. Reid, and Rolf R.~K{\"o}tter.
\newblock Connecting mean field models of neural activity to eeg and fmri data.
\newblock {\em Brain Topography}, 23(2):139--149, 2010.

\bibitem{BremaudMassoulie96}
P.~Br\'emaud and L.~Massouli\'e.
\newblock {Stability of nonlinear Hawkes processes.}
\newblock {\em Ann. Probab.}, 24(3):1563--1588, 1996.

\bibitem{bressloff}
P.~C. Bressloff.
\newblock {Stochastic neural field theory and the system-size expansion.}
\newblock {\em SIAM J. Appl. Math.}, 70(5):1488--1521, 2009.

\bibitem{Brillinger1988}
D.~Brillinger.
\newblock {Maximum likelihood analysis of spike trains of interacting nerve
  cells}.
\newblock {\em Biol. Cybern.}, 59(3):189--200, 1988.

\bibitem{cessac_discrete_time}
B.~Cessac.
\newblock A discrete time neural network model with spiking neurons: I{I}:
  Dynamics with noise.
\newblock {\em Journal of Mathematical Biology}, 62:863--900, 2011.

\bibitem{Cessac1994}
B.~Cessac, B.~Doyon, M.~Quoy, and M.~Samuelides.
\newblock Mean-field equations, bifurcation map and route to chaos in discrete
  time neural networks.
\newblock {\em Physica D: Nonlinear Phenomena}, 74:24 -- 44, 1994.

\bibitem{csiszar2006}
I.~Csisz\'ar and Z.~Talata.
\newblock Context tree estimation for not necessarily finite memory processes,
  via {BIC} and {MDL}.
\newblock {\em IEEE Trans. Inf. Theory}, 52(3):1007--1016, 2006.

\bibitem{delarue}
F.~Delarue, J.~Inglis, S.~Rubenthaler, and E.~Tanr\'e.
\newblock {Global solvability of a networked integrate-and-fire model of
  McKean-Vlasov type}, 2012.

\bibitem{dfh}
S.~Delattre, N.~Fournier, and M.~Hoffmann.
\newblock {High dimensional Hawkes processes}, 2014.

\bibitem{bresiliens}
A.~Duarte and G.~Ost.
\newblock A model for neural activity in the absence of external stimuli, 2014.

\bibitem{Ecker}
A.~S. Ecker, P.~Berens, G.~A. Keliris, M.~Bethge, N.K. Logothetis, and A.S.
  Tolias.
\newblock Decorrelated neuronal firing in cortical microcircuits.
\newblock {\em Science}, 327(5965):584--587, 2010.

\bibitem{Faugerasetal2009}
O.~{Faugeras}, J.~Touboul, and B.~Cessac.
\newblock A constructive mean-field analysis of multi-population neural
  networks with random synaptic weights and stochastic inputs.
\newblock {\em Frontiers in Comp. Neuroscience}, 3:1--28, 2009.

\bibitem{evanicolas}
N.~Fournier and E.~L\"ocherbach.
\newblock On a toy model of interacting neurons, 2014.

\bibitem{galves-locherbach-rissanen}
A.~Galves and E.~L\"ocherbach.
\newblock {\em Stochastic chains with memory of variable length.}, pages
  117--133.
\newblock TICSP Series vol. 38, 2008.

\bibitem{antonioeva13}
A.~{Galves} and E.~{L{\"o}cherbach}.
\newblock {Infinite systems of interacting chains with memory of variable
  length - a stochastic model for biological neural nets}.
\newblock {\em J. Stat. Phys.}, 151:896--921, 2013.

\bibitem{enzaantonioetal}
A.~Galves, E.~Orlandi, and D.Y. Takahashi.
\newblock Identifying interacting pairs of sites in ising models on a countable
  set., 2015.

\bibitem{Gerstner}
W.~Gerstner and W.~M. Kistler.
\newblock {\em {Spiking neuron models. Single neurons, populations,
  plasticity.}}
\newblock {Cambridge: Cambridge University Press.}, 2002.

\bibitem{hrbr}
N.~Hansen, P.~Reynaud-Bouret, and V.~Rivoirard.
\newblock {Lasso and probabilistic inequalities for multivariate point
  processes}, 2012.

\bibitem{hawkes71}
A.~G. Hawkes.
\newblock {Point spectra of some mutually exciting point processes.}
\newblock {\em J. R. Stat. Soc., Ser. B}, 33:438--443, 1971.

\bibitem{PE}
P.~Hodara and E.~L\"ocherbach.
\newblock Hawkes processes with variable length memory and an infinite number
  of components, 2014.

\bibitem{Huntsman}
M.M. Huntsman, D.M. Porcello, G.E. Homanics, T.M. DeLorey, and J.R. Huguenard.
\newblock Reciprocal inhibitory connections and network synchrony in the
  mammalian thalamus.
\newblock {\em Science}, 283:541--543, 1999.

\bibitem{Kac}
M.~Kac.
\newblock {\em {Foundations of kinetic theory.}}, pages 171--197.
\newblock Proc. 3rd Berkeley Sympos. Math. Statist. Probability 3. 1956.

\bibitem{kipnis-landim}
C.~Kipnis and C.~Landim.
\newblock {\em {Scaling limits of interacting particle systems.}}
\newblock {Grundlehren der Mathematischen Wissenschaften. 320. Berlin:
  Springer.}, 1999.

\bibitem{krumin}
M.~Krumin, I.~Reutsky, and S.~Shoham.
\newblock Correlation-based analysis and generation of multiple spike trains
  using {H}awkes models with an exogenous input.
\newblock {\em Front Comput Neurosci.}, 4:147, 2010.

\bibitem{Liggett1985}
T.M. Liggett.
\newblock {\em Interacting Particle Systems}.
\newblock Springer Berlin Heidelberg, 1985.

\bibitem{evaenza}
E.~L{\"o}cherbach and E.~Orlandi.
\newblock {Neighborhood radius estimation for variable-neighborhood random
  fields.}
\newblock {\em Stochastic Processes Appl.}, 121(9):2151--2185, 2011.

\bibitem{AAEE}
A.~De Masi, A.~Galves, E.~L{\"o}cherbach, and E.~Presutti.
\newblock {Hydrodynamic limit for interacting neurons.}
\newblock {\em J. Stat. Physics}, 0-3, 2015.

\bibitem{anna-errico91}
A.~De Masi and E.~Presutti.
\newblock {\em {Mathematical methods for hydrodynamic limits}}.
\newblock {Lecture Notes in Mathematics. 1501. Berlin: Springer-Verlag}, 1991.

\bibitem{MollerRasmussen}
J.~M{\o}ller and J.~G. Rasmussen.
\newblock {Perfect simulation of Hawkes processes.}
\newblock {\em Adv. Appl. Probab.}, 37(3):629--646, 2005.

\bibitem{Moynot02}
O.~Moynot and M.~Samuelides.
\newblock {Large deviations and mean-field theory for asymmetric random
  recurrent neural networks}.
\newblock {\em Probability Theory and Related Fields}, 123(1):41--75, 2002.

\bibitem{pinotsis}
D.~Pinotsis, P.~Robinsons, P.~beim Graben, and K.~Friston.
\newblock {Neural masses and fields: modeling the dynamics of brain activity.}
\newblock {\em Front Comput Neurosci}, 8:149, 2014.

\bibitem{rissanen}
J.~Rissanen.
\newblock A universal data compression system.
\newblock {\em IEEE Trans. Inform. Theory}, 29(5):656--664, 1983.

\bibitem{Touboul-Robert}
P.~Robert and J.~Touboul.
\newblock On the dynamics of random neuronal networks, 2014.

\bibitem{Spitzer1970}
F.~Spitzer.
\newblock Interaction of {M}arkov {P}rocesses.
\newblock {\em Adv. Math.}, 5(2):246--290, 1970.

\bibitem{sznitman}
A-S. Sznitman.
\newblock {Topics in propagation of chaos.}
\newblock {Calcul des probabilit\'es, Ec. d'\'Et\'e, Saint-Flour/Fr. 1989,
  Lect. Notes Math. 1464, 165-251 (1991).}, 1991.

\bibitem{Touboul2014}
J.~Touboul.
\newblock Propagation of chaos in neural fields.
\newblock {\em The Annals of Applied Probability}, 24(3):1298--1328, 06 2014.

\bibitem{Vreeswijk}
C.~van Vreeswijk and H.~Sompolinsky.
\newblock Chaos in neuronal networks with balanced excitatory and inhibitory
  activity.
\newblock {\em Science}, 274:1724--1726, 1996.

\end{thebibliography}

\end{document}